\documentclass[,final]
  {aipproc}

\layoutstyle{8x11double}

\begin{document}

\title{Hyperstaticity and loops in frictional granular packings}


\author{Antoinette Tordesillas}{
  address={Department of Mathematics and Statistics, University of Melbourne, 3010, Australia}
}

\author{Edward Lam}{
  address={Department of Mathematics and Statistics, University of Melbourne, 3010, Australia}
}

\author{Philip T. Metzger}{
  address={Granular Mechanics and Regolith Operations Laboratory, NASA Kennedy Space Center, Florida 32899, USA}
}

\begin{abstract}
The hyperstatic nature of granular packings of perfectly rigid disks is analyzed algebraically and through numerical simulation.  The elementary loops of grains emerge as a fundamental element in addressing hyperstaticity.  Loops consisting of an odd number of grains behave differently than those with an even number.  For odd loops, the latent stresses are exterior and are characterized by the sum of frictional forces around each loop.  For even loops, the latent stresses are interior and are characterized by the alternating sum of frictional forces around each loop.  The statistics of these two types of loop sums are found to be Gibbsian with a ``temperature'' that is linear with the friction coefficient $\mu$ when $\mu<1$.
\end{abstract}

\maketitle

\section{Analysis}

Every 2D granular packing is topologically isomorphic to a polyhedron.  The mapping replaces each grain by a vertex, each contact between grains by an edge, and each of the granular packing's pore spaces (the regions surrounded by an elementary loop of grains) by a face of the polyhedron.  A rigid boundary surrounding a granular packing is simply another node of the polyhedron located on the far side of a sphere. Thus, the polyhedron is the polygonization of a sphere and is convex, and Euler's formula implies
\begin{equation}
k=g+\lambda-\epsilon \label{CGLeps}
\end{equation}
where $k$ is the number of contacts, $g$ is the number of grains, $\lambda$ is the number of loops and $\epsilon=1$.  The value of $\epsilon$ is one less than the Euler Characteristic $\chi$ for a convex polyhedron because the boundary has not been counted as a grain, although it is counted as a vertex in Euler's formula.  With periodic boundaries around the packing, the resulting polyhedron is the polygonization of a torus and Euler's formula still applies with $\epsilon=\chi=0$.

The average number of contacts per grain (the average \textit{coordination number}) $\langle Z \rangle$, and the average number of grain per loop $\langle G \rangle$, are
\begin{eqnarray}
\langle Z \rangle = \frac{2 k}{g}, &\ \  & \langle G \rangle = \frac{\langle Z \rangle k}{\lambda} \label{ZandGdefs}
\end{eqnarray}
Some manipulation of these along with Eq.~\ref{CGLeps} obtains,
\begin{equation}
\frac{1}{\langle Z \rangle} + \frac{1}{\langle G \rangle} = \frac{1}{2} - \frac{\epsilon}{2 k} \label{ZGrel3}
\end{equation}
Dropping the last term introduces negligible error for packings with rigid boundaries if $k >> \epsilon$.

The stability equations for a quasi-static granular packings of round, frictional, 2D disks may be written as,
\begin{eqnarray}
\mathcal{A}_1 \vec{f}_n+ \mathcal{A}_2 \vec{f}_t & = & \vec{w}\\ \nonumber
\mathcal{A}_3 \vec{f}_t & = & \vec{0}
\end{eqnarray} 
where $\vec{f}_n$ is a $k$-dimensional vector consisting of all the contact normal forces in the packing, 
$\vec{f}_t$ the contact tangential forces, and $\vec{w}$ a $2g$-dimensional vector consisting of the $x$ and $y$ components of the body forces of all the grains in the packing, typically having elements 0 in the top half (representing the $x$ components) and mg in the bottom half (representing the $y$ components), where m is the grain mass and g is gravity.  The top equation is for translational stability and the bottom equation is for rotational stability. $\vec{0}$ is a $g$-dimensional null-vector because body forces do not induce torques on these disks.  Because $\vec{f}_n$ does not affect rotation, we may treat the second equation separately as if the grains are translationally frozen as in the case of a random gear network. $\mathcal{A}_3$ has dimensions $g \times k$ with $g < k$ per Eq.~\ref{ZGrel3}, so the system of rotations is underspecified (hyperstatic).  The system may be made isostatic by adding $\lambda-\epsilon$ equations, one per loop of grains to within the value of $\epsilon$.  ($\epsilon$ is related to rotation of the boundary and packing as a whole.) 

\begin{figure}
\includegraphics[angle=0,width=0.7\columnwidth,trim=0 13 0 0]{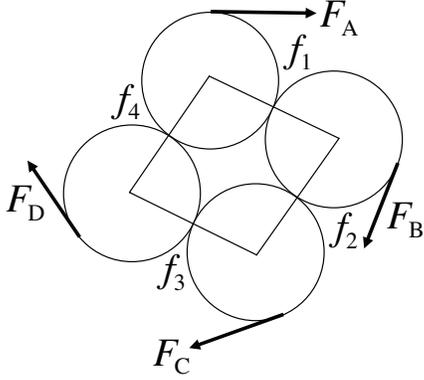}
\caption{\label{loopoffour} Example of an elementary loop. $f_i$ are interior forces. $F_\alpha$ are exterior forces.}
\end{figure}

For an elementary loop of $G$ grains (see Fig.~\ref{loopoffour} where $G=4$)
embedded in a granular packing, the rotational equations of the grains in that loop may be written as,
\begin{equation}
\mathcal{A}\vec{f}_t=-\vec{F}
\end{equation}
where the elements of the $G$-dimensional $\vec{f}_t$ are the tangential forces on the interior contacts of the loop, and the elements of $\vec{F}$ are the exterior tangential forces forces on each grain.  The determinant of this matrix,
\begin{equation}
\det\mathcal{A}=\left\{\begin{array}{lc}2&\rm{if}\ G\ \rm{odd}\\
0&\rm{if}\ G\ \rm{even}\end{array}\right.
\end{equation}
indicates that odd and even loops are fundamentally different.  If $G$ is odd, then the system of forces can be solved immediately. If $G$ is even, then the matrix $\mathcal{A}$ is singular with a null space dimensionality of one, so one of its rows must be replaced by a vector that spans the null space.  We find that vector by replacing any row with $(a_1,a_2,\dots,a_G)$ and insisting that the determinant be non-zero, and this obtains,
\begin{equation}
\sum_{n=1}^{G} (-1)^n a_n \ne 0
\end{equation}
The most symmetric treatment is to set $a_{n}=(-1)^{n}$.  This row of the new matrix calculates the alternating sum of the tangential forces around the loop, 
\begin{equation}
\beta=\sum_{n}^G(-1)^n f_n
\end{equation}
which we call the \textit{alternating loop sum}.  The value of $\beta$ is specified in the corresponding row of $F$.  A linear elasticity model of this loop can verify that $\beta$ equals the locked-in stresses in the loop, which are independent of the external forces $F_n$.

We cannot define an alternating sum around a loop when $G$ is odd, but we can calculate the non-alternating sum of interior tangential forces around the loop,
\begin{equation}
\alpha=\sum_{n=1}^G f_n
\end{equation}
for $G$ odd or even.  It turns out that
\begin{equation}
\alpha=\left\{\begin{array}{lc} -\frac{1}{2}\sum_{n=1}^G F_n & \rm{if}\ G\ \rm{odd}\\
\\
\frac{1}{2}\sum_{n=1}^G (-1)^n F_n & \rm{if}\ G\ \rm{even}\end{array}\right.
\end{equation}
These differences between odd and even loops reflect that even loops of gears are free to rotate and yet maintain a locked-in, interior stress, whereas odd loops of gears are frustrated and cannot turn and yet cannot maintain an interior stress.  Analysis of multiple loops in a granular packing shows that for each additional loop added to the packing, one more loop sum equation must be added to $\mathcal{A}_3$ to make it square and non-singular:  an alternating loop sum $\beta$ if $G$ is even, or a non-alternating loop sum $\alpha$ if $G$ is odd.  Thus, the tangential forces summed around loops (not the vector forces) are expected to be the fundamental entity in understanding hyperstaticity.

\section{Numerical Simulations}

To study the statistics of these loop sums, discrete (or distinct) element modeling (DEM) has been performed.  The DEM model comprises of a polydisperse assembly of frictional circular particles. Full details of this model are provided in \cite{tord_pm_07}.  This model has been employed to examine the constitutive response of granular assemblies, in two dimensions, under a variety of compression and penetration tests (e.g. \cite{tord_pm_07,muthuswamy_jstat_06,tord_mms_08,TorEtAl06}). The contact laws adopted are similar to other DEM simulations (e.g. \cite{luding_md_04}) which employ spring, dash-pot and friction slider to model interaction at contacts, as first proposed in \cite{IwashitaOda00}.  The key difference between this model and the classical DEM \cite{IwashitaOda00} lies in the contact moment: the model employed here incorporates a moment transfer, in accordance with the so-called modified distinct element method (MDEM) \cite{IwashitaOda00,  OdaIwashita00}. The analysis of the prior section could be extended to include moment transfer with some loss of clarity.

We performed a series of simulations involving a frictional granular packing in 2D.  In each test, a granular assembly is created from $1681$ circular particles, whose radii are chosen randomly from a uniform distribution between $0.1~mm$ and $0.15~mm$.   The particles are dropped into a box with dimensions of $10~mm\times10~mm$, under gravity, with the coefficient of friction between particles initially set to $\mu = 10^{6}$.  The assembly is then allowed to settle to a state where the kinetic energy is negligible.

All walls are assumed to have the same material properties as the particles. 
Damping coefficients are assigned according to the formulas: $b^n=0.1\sqrt{m_{min}k^n}$, $b^t=0.1\sqrt{m_{min}k^t}$,
$b^r=0.1R_{min}\sqrt{m_{min}k^r}$ where $m_{min}$ is the mass of the
smaller particle. The discrete time step used in the numerical
integration of the equations of motion is assigned a value according
to: $\Delta t=0.1\sqrt{m_{min}/k^n}$.  

The theory outlined in the previous sections addresses packings with perfect rigidity, but numerical simulations always have some compression at the contacts, which is known to affect $\langle Z \rangle$. To test the limits of the simulation and identify a range of parameters where the theory can be tested accurately, we performed fifteen simulations at a range of stiffnesses $k^r$, $k^t$ and $k^n$ and Coulomb friction coefficient $\mu$.  A summary of all the stiffness constants used in each test is presented in Table~\ref{paramsEXP}.  

\begin{table}
\caption{\label{paramsEXP} Spring stiffness values used for the normal and tangential contact force and the contact moment, $k^n$, $k^t$, $k^r$, respectively.}
\begin{tabular}{cccc}
\hline
Test & Normal $k^n$& Tangential  $k^t$ & Rotational  $k^r$ \\
 & $(N/m)$ & $(N/m)$ & $(N/m)$ \\
\hline
1 & $1.05\times10^5$ & $3.50\times10^4$ & $3.50\times10^2$ \\
2 & $2.10\times10^5$ & $7.01\times10^4$ & $7.01\times10^2$ \\
3 & $4.20\times10^5$ & $1.40\times10^5$ & $1.40\times10^3$ \\
4 & $8.41\times10^5$ & $2.80\times10^5$ & $2.80\times10^3$ \\ 
5 & $1.68\times10^6$ & $5.60\times10^5$ & $5.60\times10^3$\\
6 & $3.36\times10^6$ & $1.12\times10^6$ & $1.12\times10^4$\\
7 & $6.72\times10^6$ & $2.24\times10^6$ & $2.24\times10^4$\\
8 & $1.34\times10^7$ & $4.48\times10^6$ & $4.48\times10^4$\\
9 & $2.69\times10^7$ & $8.97\times10^6$ & $8.97\times10^4$\\
10 & $5.38\times10^7$ & $1.79\times10^7$ & $1.79\times10^5$\\
11 & $1.08\times10^8$ & $3.59\times10^7$ & $3.59\times10^5$\\ 
12 & $2.15\times10^8$ & $7.17\times10^7$ & $7.17\times10^5$\\
13 &  $8.60\times10^8$ & $2.87\times10^8$ & $2.87\times10^6$\\ 
14 & $3.44\times10^9$ & $1.15\times10^9$ & $1.15\times10^7$\\
15 & $1.38\times10^{10}$ & $4.59\times10^9$ & $4.59\times10^7$\\
\hline
\end{tabular}
\vspace{2em}
\label{tab:expPara}
\end{table}

\begin{figure}
\includegraphics[angle=0,width=\columnwidth,trim=10 20 7 12]{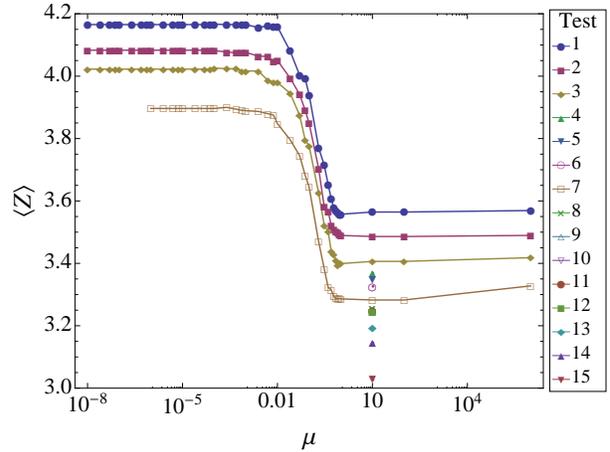}
\caption{\label{Figure:z-mu} (Colour online) $\langle Z \rangle$ as a function of Coulomb coefficient $\mu$ for several contact stiffnesses.  Additional values of $k^r$, $k^t$ and $k^n$ were tested at $\mu=10$, demonstrating that $\langle Z \rangle \to 3.0$, the isostatic value for perfect rigidity and infinite friction coefficient.}
\end{figure}

The first three tests (1-3) are performed as follows.  In each test, we lowered the value for $\mu$ and let the particles settle again to a negligible kinetic energy.  Again the value for $\mu$ is lowered and the system is left to settle. This process is repeated until $\mu$ is reduced to a value of  $10^{-9}$. The same process is used for each test, using the same values of $\mu$, as selected in test 1.   Note the amount by which $\mu$ is decreased from $10^{6}$ to $10^{-9}$ is not uniform.  Previous simulations showed that significant changes did not occur until $\mu$ is less than one.  The results for these tests are shown in Fig.~\ref{Figure:z-mu}. 

For very small values of the coefficient of friction, $10^{-9}$ to $10^{-3}$, we observe a near constant value for the average coordination number.  This value decreases with increasing particle rigidity.  The $\langle Z \rangle$ then decreases rapidly around $\mu=0.025$ to $\mu=1$, before saturating again to a near constant value.  To ensure the trends are reproducible, we repeated the test for a $k^n$ value that is two orders of magnitude higher than that used in Test 1.  As simulation times proved prohibitively long for very large values of $k^n$, test 7 was run only from $\mu=10^{6}$ down to $\mu=10^{-4}$. 

\begin{figure}
\includegraphics[angle=0,width=\columnwidth,trim=3 2 1 0]{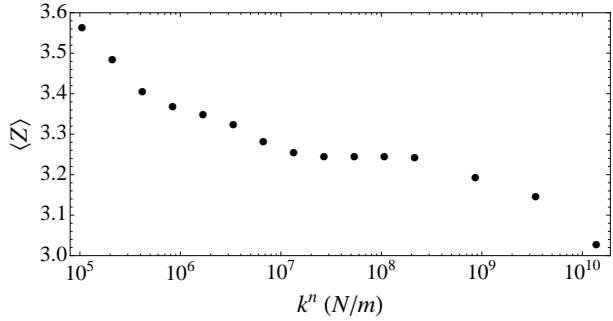}
\caption{\label{Figure:z-stiffness} $\langle Z \rangle$ as a function of stiffness $k^n$ for $\mu=10$ data points shown in Fig.~\ref{Figure:z-mu}.  $\langle Z \rangle$ approaches the isostatic limit of 3.0 as $k^n \to \infty$, however, the sudden plunge for $k^n>10^8$ is probably due to a sudden growth in the number of rattlers.}

\end{figure}

As shown in Fig.~\ref{Figure:z-mu}, we also performed an additional eleven tests (4-6, 8-15) for very high values of $k^n$ to determine the limiting value for  $\langle Z \rangle$ for perfect rigidity and infinite friction coefficient.  Fig.~\ref{Figure:z-stiffness} shows a plot of $\langle Z \rangle$ versus $k^n$ for $\mu=10$.  As the normal stiffness coefficient is increased, $\langle Z \rangle$ approaches the isostatic limit of $3$. The plunge in $\langle Z \rangle$ for  $k^n > 10^{8}$ can be attributed to an increase in the number of rattlers.


\begin{figure}
\includegraphics[angle=0,width=\columnwidth,trim=2 9 0 0]{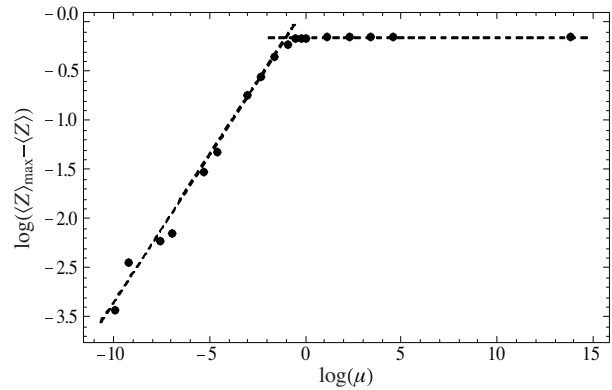}
\caption{\label{logz-logmu} $\langle Z \rangle_{\rm{max}}-\langle Z \rangle$ behaves as a power law of $\mu$ with exponent ~0.30, and transitions to a plateau near $\mu=0.1$.  The dashed lines are a guide to the eye.}
\end{figure}

As shown in Fig.~\ref{logz-logmu}, we find that the behavior of $\langle Z \rangle_{\rm{max}}-\langle Z \rangle$ is a power-law with $\mu$.

\section{Elementary Loops}

The problem of finding elementary loops is a well-studied problem in graph theory. Note that what we call loops in granular packings are called cycles in graph theory, not loops which are something different.  The problem faced is to find the minimum cycle basis of a graph. This is a set of cycles such that there is a minimal number of edges in each cycle and these cycles ``combined'' form a basis of the cycle space of the graph. We have chosen to implement an algorithm discussed in \cite{Horton87}.  The algorithm consists of four steps:   (1) find the shortest paths between every pair of vertices;  (2) generate cycles using the paths found; (3) sort all cycles by length;  (4) find all linearly independent cycles. 

\begin{figure}
\includegraphics[angle=0,width=\columnwidth,trim=1 2 4 0]{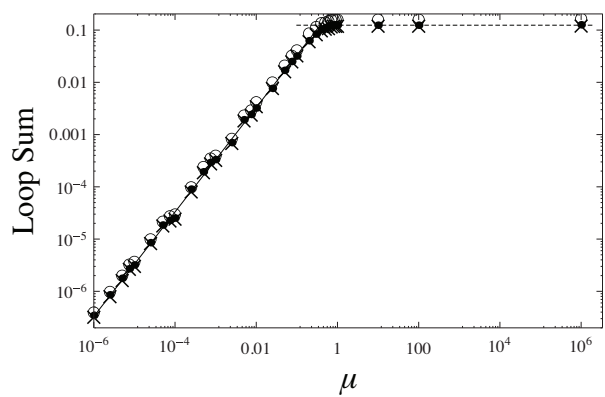}
\caption{\label{loopsums} Average value of loop sums $\langle \alpha \rangle_{\rm{odd}}$ (solid dots), $\langle \alpha \rangle_{\rm{even}}$ (X's), and average value of alternating loop sums $\langle \beta \rangle_{\rm{even}}$ (open circles) versus friction coefficient $\mu$. Solid line is a power law in $\mu$ with unity exponent.  Dashed line is a guide to the eye.}
\end{figure}

Figure~\ref{loopsums} shows the values of $\alpha$ averaged over all even loops $\langle \alpha \rangle_{\rm{even}}$, and averaged over all odd loops $\langle \alpha \rangle_{\rm{odd}}$, and the values of $\beta$ averaged over all even loops $\langle \beta \rangle_{\rm{even}}$ for all the simulations at one value of $k^n$.  

\begin{figure}
\includegraphics[angle=0,width=\columnwidth,trim=0 5 38 2.9]{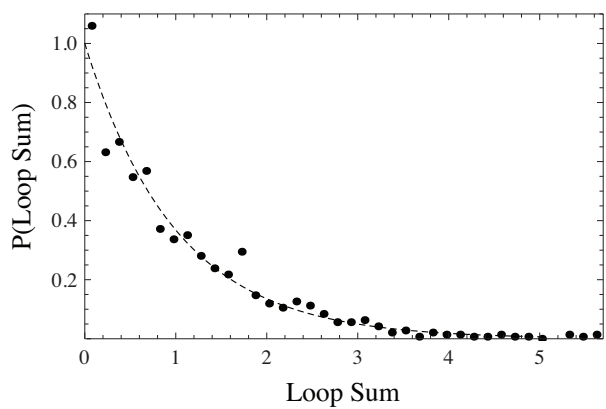}
\caption{\label{loopsumdist} Normalized distribution of loop sums over all odd loops in a packing with $\mu = 10^{-5}$.  Dashed line is an exponential decay.}
\end{figure}

\begin{figure}
\includegraphics[angle=0,width=\columnwidth,trim=3.5 9 3.7 10]{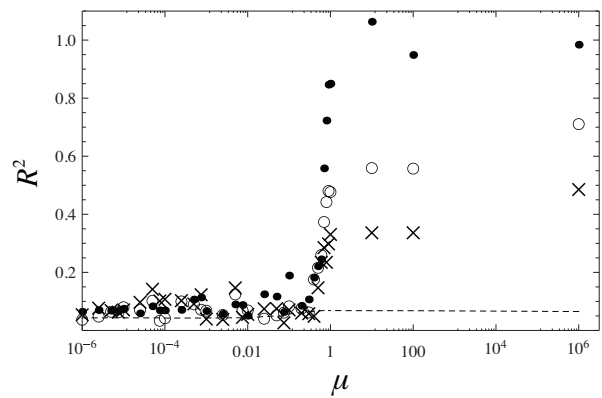}
\caption{\label{resids} $R^2$ (sum of squares) norm to quantify how closely the distribution of $\alpha_{\rm{odd}}$ (solid dots), $\alpha_{\rm{even}}$ (X's), and $\beta_{\rm{even}}$ (open circles) follows the Gibbs distribution for different values of $\mu$.  Dashed line is the expectation value for a perfect Gibbs distribution sampled as herein.}
\end{figure}

In all cases the distribution of loop sums appears to be an exponential decay with the decay constant equal to the inverse of the average value of the loop sums.  In other words, it appears to be a Gibbs distribution.  An example for $\mu = 10^{-5}$ is shown in Fig.~\ref{loopsumdist}. 
To test how closely these distributions follow a pure exponential decay, the least-squares difference $R^2$ is calculated and summed over all bins in the distribution.  In Fig.~\ref{resids} $R^2$ is plotted for each of the three types of loop sums and for packings at each value of $\mu$.
It is found that the distributions become pure Gibbsian for $\mu<1$, which is the same place where $\langle \alpha \rangle$ and $\langle \beta \rangle$ become proportional to $\mu$ (cf. Fig.~\ref{loopsums}).  $\mu$ (or a linear function of it) may be interpreted as the latent stress temperature in a packing below that limit.

\bibliographystyle{aipproc}

\end{document}